\documentclass[twocolumn]{aastex61}
\usepackage{graphics,graphicx,amssymb}

\begin{document}

\title{The High Frequency Radio Emission of the Galactic Center
  Magnetar SGR J1745$-$29 During a Transitional Period}

\author[0000-0003-4679-1058]{Joseph D. Gelfand}
\affiliation{NYU Abu Dhabi, United Arab Emirates}
\affiliation{NYU Center for Cosmology and Particle Physics, New York,
  NY 10003, USA }
\author[0000-0001-5799-9714]{Scott Ransom}
\affiliation{NRAO}
\author[0000-0003-1443-593X]{Chryssa Kouveliotou}
\affiliation{The George Washington University, USA}
\affiliation{Astronomy, Physics and Statistics Institute of Sciences
  (APSIS), Washington, DC 20052, USA}
\author[0000-0001-8530-8941]{Jonathan Granot}
\affiliation{The Open University of Israel}, Israel 
\author{Alexander J. van der Horst}
\affiliation{The George Washington University, USA} 
\affiliation{Astronomy, Physics and Statistics Institute of Sciences
  (APSIS), Washington, DC 20052, USA} 
\author[0000-0001-8630-5435]{Guobao Zhang} 
\affiliation{NYU Abu Dhabi, United Arab Emirates}
\affiliation{Yunnan Observatories, Chinese Academy of Sciences, 396
  Yangfangwang, Guandu District, Kunming 650216, China} 
\affiliation{Key Laboratory for the Structure and Evolution of
  Celestial Objects, Chinese Academy of Sciences, 396 Yangfangwang,
  Guandu District, Kunming 650216, China} 
\affiliation{Center for Astronomical Mega-Science, Chinese Academy of
  Sciences, 20A Datun Road, Chaoyang District, Beijing 100012, China} 
\author[0000-0002-5274-6790]{Ersin G{\"O}{\u G}{\"U}{\c S}}
\affiliation{Sabanc\i~University, Turkey}
\author[0000-0002-9396-9720]{Mallory S. E. Roberts}
\affiliation{NYU Abu Dhabi, United Arab Emirates}
\author[0000-0002-4187-4981]{Hend Al Ali}
\affil{NYU Abu Dhabi, United Arab Emirates}

\begin{abstract}
  The origin of the high-frequency radio emission detected from
  several magnetars is poorly understood. In this paper, we report the
  $\sim40~{\rm GHz}$ properties of SGR J1745$-$29 as measured using
  Jansky Very Large Array (JVLA) and Robert C. Byrd Green Bank
  Telescope (GBT) observations between 2013 October 26 and 2014 May
  31.  Our analysis of a Q-band (45 GHz) GBT observation on 2014 April
  10 resulted in the earliest detection of pulsed radio emission at
  high frequencies ($\gtrsim20~{\rm GHz}$); we found that the average
  pulse has a singly peaked profile with width $\sim 75~{\rm ms}$
  ($\sim2\%$ of the $3.764~{\rm s}$ pulse period) and an average
  pulsed flux density of $\sim100~{\rm mJy}$. We also detected very
  bright, short $(<10~{\rm ms})$ single pulses during $\sim70\%$ of
  this neutron star's rotations, and the peak flux densities of these
  bright pulses follow the same log-normal distribution as measured at
  8.5~GHz.  Additionally, our analysis of contemporaneous JVLA
  observations suggest that its 41/44 GHz flux density varied between
  $\sim1-4$~mJy during this period, with a $\sim2\times$ change
  observed on $\sim20$~minute timescales during a JVLA observation on
  2014 May 10.  Such a drastic change over short time-scales is
  inconsistent with the radio emission resulting from a shock powered by
  the magnetar's supersonic motion through the surrounding medium, and
  instead is dominated by pulsed emission generated in its
  magnetosphere.
  \end{abstract}

\section{Introduction}
\label{intro}

Magnetars are believed to be young, isolated neutron stars with
extremely strong surface and internal magnetic fields.  The resultant
stresses are thought to twist the external magnetic field
\citep{thompson00}, generating persistent currents in the
magnetosphere (e.g., \citealt{thompson08, beloborodov13}).  The
eventual untwisting of the magnetar's external magnetic field is
thought to (e.g., \citealt{beloborodov09}) result in an ``activation''
event where the source X-ray flux rapidly increases by orders of
magnitude, after which it decays over weeks to months to a new
quiescent level (e.g., \citealt{ibrahim04, mori13}).  Such events have
now been observed from about a dozen magnetars, and in a few sources
are contemporaneous with the onset of pulsed radio emission (e.g.,
\citealt{camilo06, camilo07c, rea13}), likely a result from the
ensuing magnetospheric currents. However, the detection of radio
pulsations from a magnetar in X-ray quiescence \citep{levin10}
suggests that these currents can persist for a long time, though the
cessation of radio pulsations does not appear to be connected to the
source behavior at X-ray energies \citep{camilo16}.

Perhaps not surprisingly, the properties of the pulsed radio emission
detected from magnetars is significantly different from that observed
from ``normal'' radio pulsars.  Studies of the first radio-emitting
magnetar, XTE J1810$-$197, indicated its radio pulse profile, pulsed
flux density, and pulsed spectrum all varied significantly over
timescales as short as a few hours or days (e.g.,
\citealt{camilo07b,camilo07d}).  The detection of similar behavior
\citep{camilo08} from 1E~1547.0$-$5408 \citep{camilo07c}
suggested such variability is commonplace among magnetars -- in stark
contrast to the majority of radio pulsars whose average pulse profile
remains constant for years.  Additionally, the pulsed radio emission
from magnetars typically has a flat (spectral index $\alpha \sim 0$,
where flux density $S_\nu \propto \nu^\alpha$) or curved (peak
$\nu\sim10~{\rm GHz}$) spectrum (e.g., \citealt{camilo08, kijak13})
observed between $\sim1-100~{\rm GHz}$ from magnetars -- in sharp
contrast with the very steep (average $\alpha \sim -1.6$) spectrum
typically observed from most radio pulsars (e.g.,
\citealt{lorimer12}).  Furthermore, magnetars emit single, extremely
bright and short (few millisecond) long radio pulses far more often
than ``normal'' radio pulsars (e.g., \citealt{serylak09, levin12}).
While the origins of these differences -- in particular, why the
pulsed radio spectrum of magnetars extends to much higher frequencies
than that of ``normal'' radio pulsars are not yet known, it suggests
the leptons responsible for the pulsed radio emission in magnetars
have a different origin and/or acceleration mechanism than those
responsible for the pulsed radio emission from ``normal'' pulsars.
Possibilities are that, in magnetars, these particles are created when
X-ray photons emitted from the surface interact with $\gamma$-ray
photons generated in the magnetosphere (e.g., \citealt{thompson08,
  thompson08b}) or are accelerated by currents powered by the
untwisting of the magnetic field lines (e.g., \citealt{beloborodov09,
  beloborodov13}).

SGR J1745$-$29 was discovered due to a rapid, significant increase in
its X-ray flux on 2013 April 24. Analysis of an observation $\sim4-5$
days later detected pulsed radio emission (e.g., \citealt{rea13}),
whose properties -- particularly at high ($\gtrsim10$~{\rm GHz})
frequencies -- have changed considerably since this initial detection.
Analysis of Australia Telescope Compact Array (ATCA) observations on
2013 May 1 and 2013 May 31 indicated that the $16-20~{\rm GHz}$ pulsed
radio emission of this source had a fairly steep spectrum ($\alpha
\sim -2$; \citep{shannon13}) -- suggesting very faint ($<1~{\rm mJy}$)
emission at higher frequencies.  However, analysis of a Karl G. Jansky
Very Large Array (JVLA) observation on 2014 Feb 21 measured a 41~GHz
flux density of $1.62\pm0.02$~mJy \citep{yusef14} -- $\sim40\times$
the expected value based on measurements of the pulsed radio spectrum
on 2013 May by \citet{shannon13}.  This initial detection of SGR
J1745$-$29 at 41~GHz was contemporaneous with significant changes in
the 8.5 GHz pulsed flux density and pulse profile \citep{lynch14},
possibly suggesting that the appearance of high-frequency radio
pulsations is related to its behavior at lower frequencies.
Subsequent observations detected pulsed radio emission from this
magnetar at frequencies as high as 225~GHz \citep{torne15}.

In this paper, we report the results of additional $\sim40~{\rm GHz}$
observations of this magnetar between 2013 October and 2014 May, the
period when high-frequency pulsed radio emission was first detected
and the properties of its lower frequency pulsed radio emission
changed significantly.  In \S\ref{gbt}, we present the results of a
45~GHz Robert C. Byrd Green Bank Telescope (GBT) observation on 2014
April 10, which resulted in the earliest detection of $>20~{\rm GHz}$
pulsations from this magnetar.  In \S\ref{jvla}, we present our
analysis of 41/44~GHz JVLA observations between 2013 October 26 and
2014 May 31, during which we measured significant changes in flux
density on both short ($\sim20$ minutes) and long (weeks) timescales.
In \S\ref{conclusion}, we summarize our results.

\section{Green Bank Radio Telescope}
\label{gbt}

We observed SGR J1745$-$29 with the GBT for two hours, starting on
2014 April 10 08:30 (UT) in Q band (central frequency of 45~GHz).
During this observation, the system temperature was $T_{\rm
  sys}\approx 80~{\rm K}$, the zenith opacity was $\tau_{\rm zenith}
\approx 0.16$, and the $21^\circ-22^\circ$ elevation resulted in an
airmass ${\rm sec}(z)=2.7$.  However, due to instrumental
difficulties, we obtained only $\sim35$ minutes of good data with the
Green Bank Ultimate Pulsar Processing Instrument (GUPPI;
\citealt{duplain08}), with an 800~MHz bandwidth, and about $\sim30$
minutes of good data with the Versatile GBT Astronomical Spectrometer
(VEGAS; \citealt{bussa12}), with a (usable) bandwidth of 5.4~GHz with
1~ms sampling.  We present the results of the more sensitive VEGAS
dataset below; we also searched for pulsations in the less sensitive
GUPPI dataset but this effort yielded no detection.  Unfortunately,
VEGAS was not able to record the data needed to measure the
polarization of this radio emission..  We used the {\tt rednoise}
routine \citep{lazarus15} in PRESTO \citep{ransom01} to remove, in the
frequency domain, the quasi-periodic oscillations introduced by
atmospheric variability.

\begin{figure}[tbh]
  \begin{center}
    \includegraphics[width=0.45\textwidth]{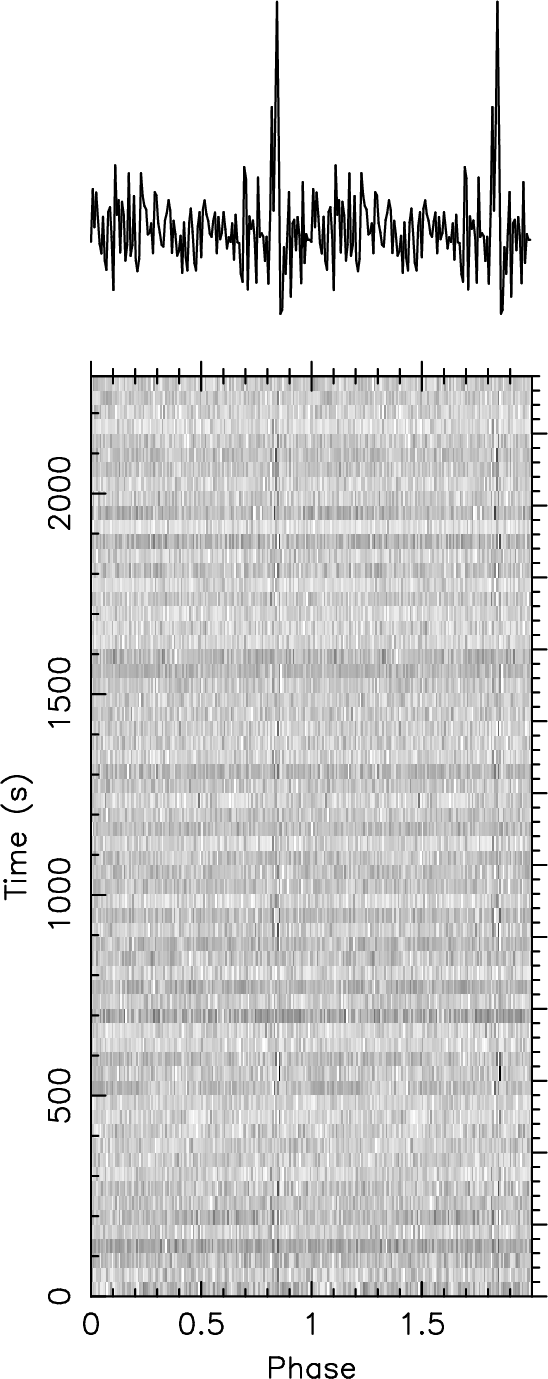}   
  \end{center}
  \vspace*{-0.5cm}
  \caption{{\it Top:}The 45~GHz pulse profile of SGR J1745$-$29 after
    folding the data for the best fit period of $P=3.763504~{\rm
      s}$. {\it Bottom}: Amplitude (greyscale, with darker grey
    indicating a higher amplitude) as function of pulsar phase and
    time during this observation.}
  \label{fig:profile}
\end{figure}

We then used PRESTO to search for pulsations in the de-reddened
time-series.  This analysis indicated statistically significant
pulsations with a period $P=3.763504~{\rm s}$ consistent with previous
measurements (period $P\approx3.764~{\rm s}$; e.g, \citealt{mori13}),
with a singly peaked radio pulse of $\sim75~{\rm ms}$ duration,
(considerably longer that the $\sim1~{\rm ms}$ time resolution of
VEGAS), roughly $\sim2\%$ of the pulse period (Figure
\ref{fig:profile}).  This is considerably different than the
integrated pulse profile detected at 8.5~GHz before (e.g.,
\citealt{lynch14b}) and after (e.g., \citealt{torne15, yan15}) this
GBT observation, but comparable to that measured at higher frequencies
($\gtrsim87~{\rm GHz}$) in the following months (e.g.,
\citealt{torne15}).  The similarity between the width of the 45~GHz
pulse and the ``third'' component in the 8.7 GHz pulse profile which
appeared in 2014 Jan -- March \citep{lynch14b} suggests the two are
possibly related.  Unfortunately, the lack of absolute phase
information prevents us from making a stronger connection between
these two features.

Since, at this frequency and LST (local sidereal time), no test pulsar
suitable for flux calibration was observable, we estimated the pulsed
flux density using the pulsar radiometer equation (Appendix A1.4 in
\citealt{lorimer12}). For the values of $T_{\rm sys}$, $\tau_{\rm
  zenith}$, and airmass given above, and the known GBT Q-band gain of
$0.68~{\rm K/Jy}$, we derive an average pulsed flux density of
$\sim100\mu$Jy.  This value corresponds to the average fluence of a
single pulsed distributed over the magnetar's entire rotation period,
with the pulsed emission from this magnetar having a much higher peak
brightness.

Furthermore, we found $\gtrsim430$ time bins where the measured
signal-to-noise ratio was $>5$, a significantly higher number than
expected assuming Gaussian noise.  Their intensity suggests that they
are radio frequency interference (RFI) or short bursts of radio
emission as observed from other magnetars (e.g., \citealt{camilo07b}).
Typically, one would distinguish these two possibilities by looking
for dispersion -- the change in the arrival time of photons with
different frequencies due to their propagation through an intervening
plasma.  Since RFI is generated locally (either on the Earth's surface
or by orbiting satellites), such signals typically are not dispersed,
while pulsed emission from astronomical objects is.  Unfortunately,
despite the very large dispersion measure (DM) towards SGR J1745$-$29
(${\rm DM = 1650\pm50 cm^{-3}~pc}$; \citealt{shannon13}), this effect
is immeasurably small at 45~GHz.  An additional method for
distinguishing between astronomical pulses and RFI is too look for
Faraday rotation, the change in polarization angle with frequency, but
this also was not possible due to the lack of polarization information
recorded by the VEGAS spectrometer.

\begin{figure*}[tbh]
  \begin{center}
    \includegraphics[width=0.475\textwidth]{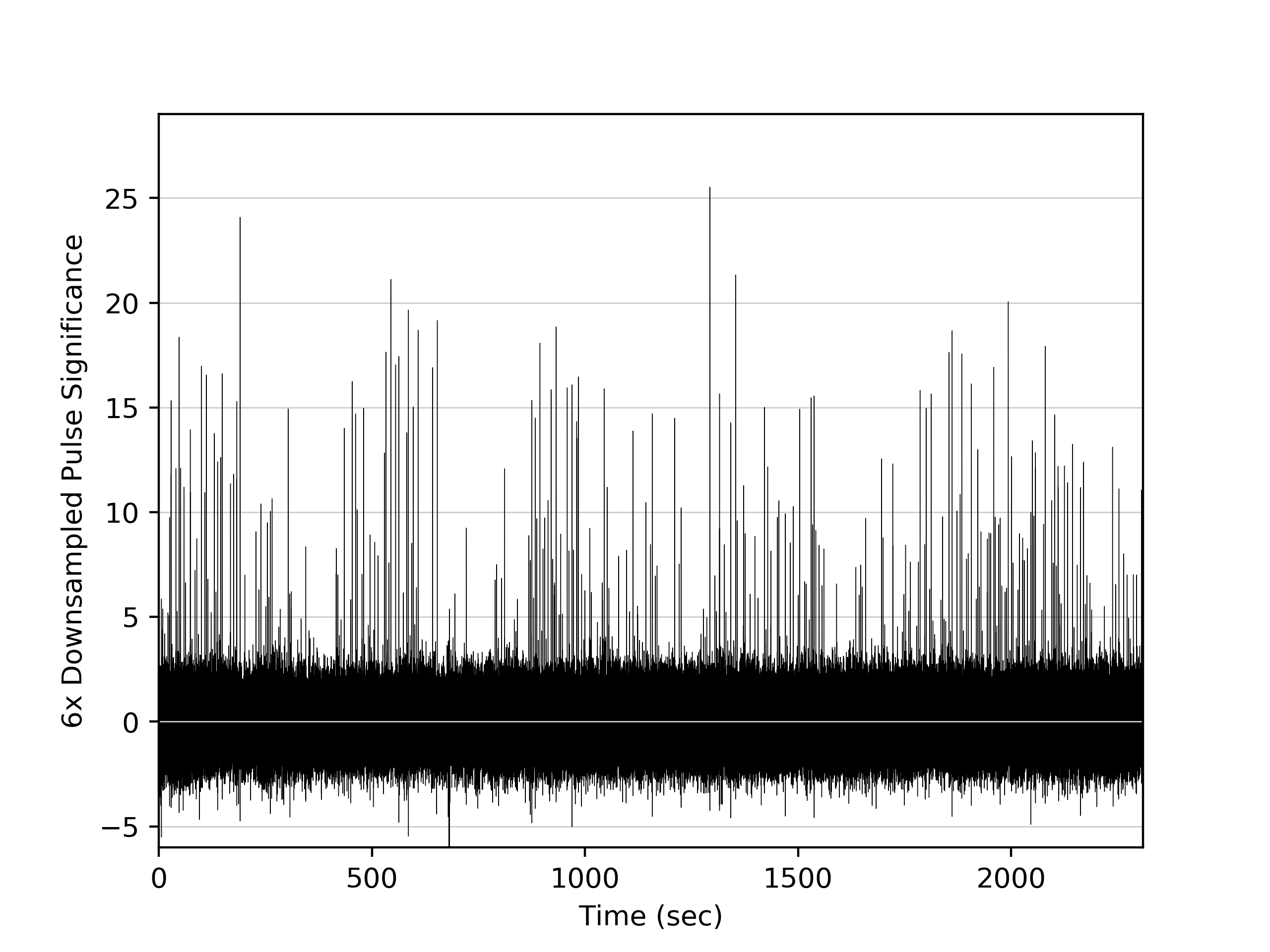}
    \includegraphics[width=0.475\textwidth]{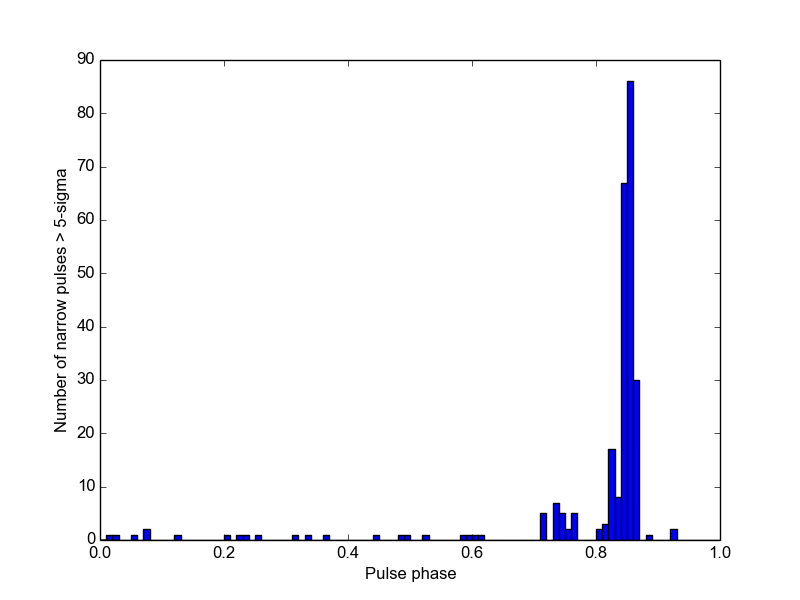} 
  \end{center}
  \caption{{\it Left}: Signficance of pulses detected over a 6~ms time
    period (the average length of the single pulses discussed in the
    text) during the 45 GHz GBT observation described in \S\ref{gbt}.
    This time period is 6$\times$ the 1~ms time resolution of VEGAS,
    and is equivalent to downsampling the recorded data by $6\times$.
    {\it Right}: Pulse phase resulting from folding the arrival times
    of samples with S/N $>$ 5 with the observed period of the
    magnetar.}
  \label{fig:voltages}
\end{figure*}

Instead, we folded the arrival time of these pulses with the
magnetar's rotational period. Since RFI is uncorrelated with magnetar
activity, it should be evenly distributed across all rotational phases
-- in contrast with pulses produced in the magnetosphere.  Indeed,
after folding, we found the arrival times of these samples are heavily
concentrated between a pulse phase of $\sim0.8-0.9$ (Figure
\ref{fig:voltages}), a clear indication they are emitted by the
magnetar.  These pulses had an average width of $\sim4.62~{\rm ms}$
($\sim0.1\%$ of the rotational period) -- comparable to similar pulses
observed from other magnetars (e.g, \citealt{camilo07d, levin12}), and
considerably shorter than the $\sim75~{\rm ms}$ width ($\sim2\%$ of
the rotational period) of the average pulse.  However, the similarity
between the average pulse profile (Figure \ref{fig:profile}) and the
phase distribution of the bright radio pulses (Figure
\ref{fig:voltages}) suggests the ``average'' pulsed radio emission
from this magnetar may be a collection of single, bright pulses which
vary in pulse phase and intensity -- similar to what is believed to
occur in other magnetars (e.g. \citealt{levin12}) and the Crab pulsar
at low radio frequencies (e.g., \citealt{karuppusamy10} and references
therein).  In fact, these single bright pulses dominate the fluence
from the pulsar, and together they account for the measured pulse flux
density and duration derived from the timing analysis above.  This
suggests the pulsed radio emission from the magnetar is dominated by
the sporadic generation of bright, short bursts.

Thanks to the multitude of single 45~GHz pulses detected during this
GBT observation, we can compare their properties with those detected
from this magnetar at $\sim8.5~{\rm GHz}$.  First the detection of 434
bright ($>5 \sigma$) pulses during our GBT observation, which only
lasted $\sim610$ pulse periods, implies that this magnetar produces
such pulses in $\sim70\%$ of neutron star rotations.  This fraction is
significantly higher than the $\sim3\%$ (53 out of 1913) derived from
8.7~GHz observations several months later (between 2014 June --
October) \citep{yan15}.  This discrepancy is unlikely to result from
using a different criterion to select ``bright'' pulses since applying
the criterion used by \citet{yan15} -- peak fluxes $>10\times$ the
peak flux of the integrated pulse profile -- results in a nearly
identical collection of pulses.  This large difference suggests that
either bright pulses are more common at higher frequencies, or the
rate of bright pulses can vary significantly with time.

\begin{figure}[tbh]
  \begin{center}
    \includegraphics[width=0.475\textwidth]{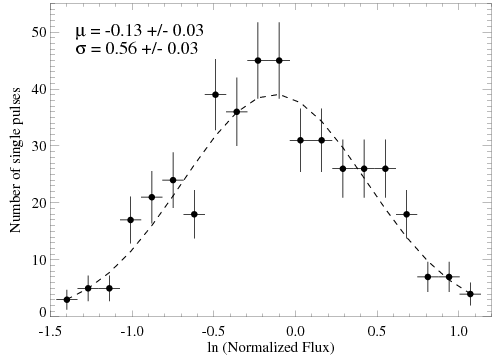}
  \end{center}
  \caption{Normalized flux distribution of 45 GHz single pulses
    detected from SGR 1745$-$29, overlaid with the best fit log-normal
    distribution as defined in Equation \ref{eqn:lognorm}.  The
    1$\sigma$ error bars correspond to the square-root of the number
    of single pulses in each log normalized flux bin.}
  \label{fig:single_fluxdist}
\end{figure}

We also compared the flux distribution of these 45~GHz single pulses
to those detected at 8.7/8.6~GHz from this magnetar \citep{lynch14b,
  yan15}.  Unlike ``normal'' radio pulsars, for which the flux
distribution of giant pulses is well described by a power-law (e.g.,
\citealt{karuppusamy10}), the peak flux of single bright pulses from
magnetars, like that of ``regular'' pulses from ``normal'' radio
pulsars (e.g., \citealt{burke-spolaor12}), is consistent with a
log-normal distribution (e.g., \citealt{levin12}):
\begin{eqnarray}
  \label{eqn:lognorm}
  N(x_{\rm min}<x<x_{\rm max}) & = & Ce^{-\frac{(x-\mu)^2}{2\sigma^2}}
\end{eqnarray}
where $x=\ln\left(\frac{S_\nu}{\langle S_\nu \rangle} \right)$,
$S_\nu$ is the peak flux density of an individual single pulse, and
$\langle S_\nu \rangle$ is the average peak flux density of all the
single pulses ($\langle S_\nu \rangle = 641~{\rm mJy}$ for the 45~GHz
single pulses detected in our GBT observation).  As shown in Figure
\ref{fig:single_fluxdist}, this function (Equation \ref{eqn:lognorm})
reproduces the measured distribution of peak flux densities for $\mu =
-0.13\pm0.03$ and $\sigma=0.56\pm0.03$.  This value of $\sigma$ is
similar to that measured at 8.6/8.7 GHz both before ($\sigma=0.49$,
\citealt{lynch14b}) and after ($\sigma = 0.57\pm 0.02$;
\citealt{yan15}) our GBT observation -- possibly suggesting a common
generation mechanism.

\section{Jansky Very Large Array}
\label{jvla}

SGR J1745$-$29 was serendipitously observed by the JVLA in its A- and
B-configurations as part of NRAO's recent service monitoring campaign
of Sgr A$^{\star}$ (project code TOBS00006) \citet{chandler13a,
  chandler13b, chandler14a, chandler14b}.  During these observations,
the WIDAR correlator was configured to achieve a 2~GHz bandwidth
centered at 41~GHz.  In the analysis described below, we used the
calibrated measurement sets available on the NRAO Archive, where flux
densities were calibrated using short observations of 3C286, and the
bandpass was calibrated using short observations of NRAO~530 and
3C286.  At 41~{\rm GHz}, Sgr A$^{\star}$ (and SGR J1745$-$29) were
only observed for $\sim6$ minutes on source, and the phases were
determined by self-calibrating on Sgr~A$^\star$ in the center of the
field\footnote{See {\tt
    https://science.nrao.edu/science/science-observing} for additional
  details.}.

The radio environment around Sgr A$^\star$ is extremely complicated;
it includes multiple diffuse radio sources which are most likely H{\sc
  ii} regions and stellar wind bubbles powered by the numerous massive
stars in this region (e.g., \citealt{zhao98}).  To minimize the impact
of this diffuse emission on our measurements of SGR J1745$-$29, we
only used baselines $>500~{\rm k}\lambda$ ($>4.8~{\rm km}$) in length
to produce our radio images of this field -- effectively removing all
sources $>0\farcs4$ in size.  We first imaged the remaining data using
the {\sc casa} task {\tt clean}, weighting the visibilities on
different $u-v$ baselines using the ``Briggs'' function
\citep{briggs95} with a ``robust'' parameter of 0.5.  The resultant
image was then deconvolved using the {\sc casa} task {\tt clean}, with
``CLEAN boxes'' being interactively placed around Sgr A$^\star$,
magnetar SGR J1745$-$29, and any other sources of emission which
appeared after each deconvolution cycle.  We measured the flux density
of SGR 1745$-$29 using the {\sc miriad} \citep{sault95} task {\tt
  imfit} to fit a point source at the magnetar's location in the
$>500~{\rm k}\lambda$ image.  The presence of the much brighter Sgr
A$^{\star}$ prevented us from measuring the magnetar's flux density by
modeling the measured $u-v$ visibilities with a point source at its
location.

As shown in Table \ref{tab:jvlaflux}, eliminating data from shorter
baselines generally increased the significance of the magnetar's
detection.  The flux densities estimated using the method described
above typically have substantially larger error bars than those
previously reported by \citet{yusef15}.  Using the larger error bars
obtained from our analysis, we find values which are in general
consistent with but lower than, those obtained by those authors.  The
main difference between our two analyses is that \cite{yusef15} do not
filter out the shorter baselines and modeled the emission from SGR
J1745$-$29 with a 2D Gaussian in the image plane \citep{yusef15}.
Both differences increase the possible contamination from unrelated
diffuse emission around this source, and explain why these authors
measure a higher, more precise flux density for SGR 1745$-$29.

\begin{table*}[tbh]
  \begin{center}
    \small
    \begin{tabular}{ccccc}
      \hline
      \hline
      Obs. Date & JVLA  & \multicolumn{3}{c}{SGR J1745$-$29 41 GHz
        Flux Density [mJy] } \\
      YYYY MMM DD & Config. & $>500$~k$\lambda$ & All data &
      \citet{yusef15}\\ 
      \hline
      2013 Oct 26 & B & $2.0\pm1.0$ & $0.50\pm0.36$ & $<0.82$ \\
      2013 Nov 29 & B & $1.0\pm0.6$ & $0.83\pm0.46$ & $<0.70$ \\ 
      2013 Dec 29 & B & $1.2\pm0.7$ & $1.2\pm1.5$ & $<1.52$\\
      2014 Feb 15 & A & $2.1\pm0.4$ & $1.6\pm0.3$ & $1.85\pm0.07$\\ 
      {\it 2014 Feb 21} & A & $\cdots$ & $1.62\pm0.02$ & $1.62\pm0.04$ \\
      2014 Mar 22 & A & $2.1 \pm 0.3$ & $0.86\pm0.22$  & $1.24\pm0.02$\\
      %% {\bf 2014 Apr 10} & {\bf GBT} & \multicolumn{2}{c}{$\sim0.055$} &
      %% $\cdots$ \\
      2014 Apr 26 & A & $0.91\pm0.30$ & $0.52\pm0.25$ & $1.20\pm0.07$ \\
      2014 May 10 & A & $1.21\pm0.22$ & $1.10\pm0.21$ & $\cdots$  \\  
      2014 May 31 & A & $3.5\pm0.4$ & $3.5\pm0.4$ & $2.94\pm0.12$ \\
      \hline
      \hline
    \end{tabular}
  \end{center}
  \caption{Compilation of measured values of the 41 GHz flux density
    of SGR J1745$-$29 between 2013 October and 2014 May (the reported
    flux density on 2014 May 10 is actually at 44~GHz.).  The reported
    flux density on 2014 Feb 21 is taken from \citet{yusef14}, while
    the rest were derived using the procedures described in
    \S\ref{jvla}.  The upper-limits quoted by \citet{yusef15}
    correspond to the flux density required for a $3\sigma$
    detection.}
  \label{tab:jvlaflux}
\end{table*}

We also analyzed an additional JVLA observation of this field taken on
2014 May 10 (Project AG941), which observed SGR J1745$-$29 for
$\sim90$ minutes with the WIDAR correlator in 8-bit mode in order to
achieve the maximum $\sim2~{\rm GHz}$ bandwidth centered at 44~GHz --
slightly different than the 41~GHz JVLA observations described above
and the 45~GHz GBT observation discussed in \S\ref{gbt}.  These data
were analyzed using CASA v4.3.1 \citep{casa}, with the flux density
scale, antenna delays, gains, and bandpass all calibrated using
observations of 3C~286, while the phases were determined using
self-calibration since Sgr A$^\star$ was again in the field.  The data
were also imaged using the {\sc casa} task {\tt clean} for a
``Briggs'' weighting \citep{briggs95} of the different $u-v$ baselines
with a ``robust'' parameter of 0.5.  The resultant image was then
deconvolved, again using the {\sc casa} task {\tt clean}, with ``CLEAN
boxes'' being interactively placed around Sgr A$^\star$, magnetar SGR
J1745$-$29, and any other sources of emission revealed by further
deconvolution cycles.  The flux density of both Sgr A$^{\star}$ and
SGR J1745$-$29 were then determined by fitting the resultant image
with a point at their locations using the {\sc miriad} task {\tt
  imfit}.  Sgr A$^{\star}$ has a 44~GHz flux density of $1.59~{\rm
  Jy}$ in data collected on all baselines as well as in data collected
only in baselines $>500~{\rm k}\lambda$ in length -- consistent with
other measurements around this date \citep{yusef15}.

\begin{table*}[tbh]
  \begin{center}
    \begin{tabular}{ccccccc}
      \hline
      \hline
      Source & \multicolumn{4}{c}{Time Range [UTC]}\\
      & 08:23:00$-$08:41:30 & 08:41:30$-$09:03:00 &
      09:05:00$-$09:24:00 & 09:24:00$-$09:43:00 \\
      \hline
      SGR J1745$-$29 & $1.99\pm0.38$~mJy & $0.88\pm0.47$~mJy &
      $1.31\pm0.50$~mJy & $0.86\pm0.52$~mJy \\  
      Sgr A$^\star$ & 1.590~Jy & 1.585~Jy & 1.590~Jy & 1.588~Jy  \\
      Image rms & 0.52~$\mu$Jy~beam$^{-1}$ & 0.59~$\mu$Jy~beam$^{-1}$
      & 0.52~$\mu$Jy~beam$^{-1}$ & 0.50~$\mu$Jy~beam$^{-1}$ \\
      \hline
      \hline
    \end{tabular}
  \end{center}
  \caption{44 GHz flux density of SGR 1745$-$29 and Sgr A$^{\star}$ as
    measured on 2014 May 10 in mJy.  The error on the flux density of
    Sgr A$^\star$ in each time period is $\sim0.6~{\rm mJy}$.}
  \label{tab:ag941fluxes}
\end{table*}

\begin{figure*}
\begin{center}
  \includegraphics[width=0.225\textwidth]{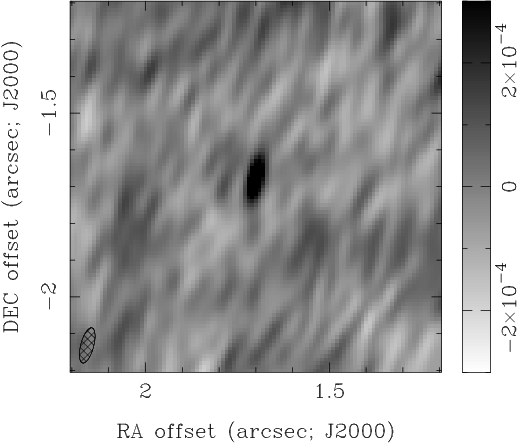}
  \includegraphics[width=0.225\textwidth]{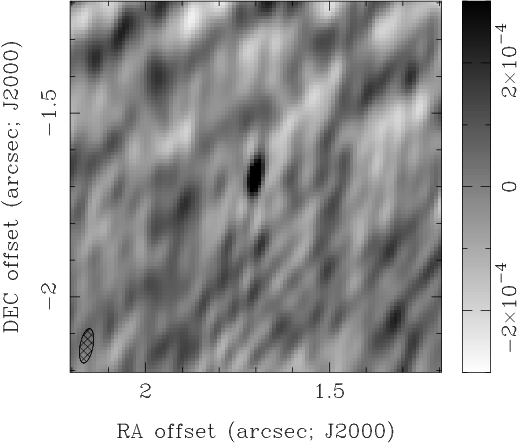}
  \includegraphics[width=0.225\textwidth]{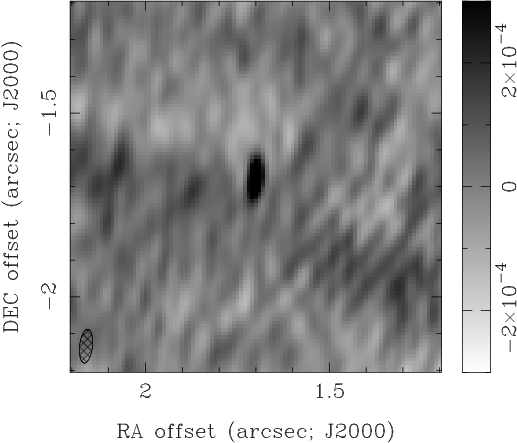}
  \includegraphics[width=0.225\textwidth]{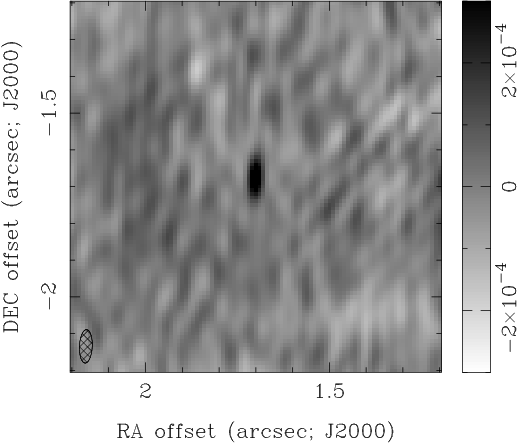}
\end{center}
\caption{44 GHz images centered on the position of SGR J1745$-$29
  produced using data from the four time periods listed in Table
  \ref{tab:ag941fluxes}.  In each image, the grey scale ranges from
  $-0.3~{\rm mJy~beam}^{-1}$ to $0.3~{\rm mJy~beam}^{-1}$, and the
  size and orientation of the resolving beam is shown in the bottom
  left corner.}
\label{fig:snapshots}
\end{figure*}

We searched for changes in the magnetar's flux density during this
observation by dividing this dataset into $\sim20$ minute increments,
imaging each time period separately using the same procedure as
described above (Figure \ref{fig:snapshots}) and again measuring both
the flux density of SGR J1745$-$29 and Sgr A$^\star$ by fitting the
resultant image with a point at its location using the {\sc miriad}
task {\tt imfit}.  As listed in Table \ref{tab:ag941fluxes}, the flux
density of the magnetar appeared to vary by a factor of $\sim2$
between successive $\sim20$ minute periods in this observation.  The
constant flux density measured for Sgr A$^{\star}$ during this
observation, as well as the nearly constant noise level in the image
(Table \ref{tab:ag941fluxes} as measured by calculating the rms of
pixels in a large source-free region near the magnetar using the {\sc
  karma} tool {\tt kvis} \citep{gooch11}) suggests this variability is
not an artifact of either our calibration or imaging technique.

The large changes in the magnetar's radio flux density on such short
timescale allows us to test the suggestion of \citet{yusef15} that the
observed radio flux is primarily unpulsed emission generated by the
interaction between the magnetar's rotation powered wind and
surrounding medium.  These authors argue that, if correct, the
unpulsed flux density is proportional to the magnetar wind's ram
pressure $P_{\rm ram} = \rho v_{\rm rel}^2$, where $\rho$ is the
density of the surrounding medium and $v_{\rm rel}$ is the relative
speed of the magnetar to its surrounding \citep{yusef15}.  This
suggests the factor of $\sim2$ change in flux observed during the 2014
May 10 JVLA observation (Table \ref{tab:ag941fluxes}) results from a
$\sim2\times$ change in ambient density and/or a $\sqrt{2}\times$
change in $v_{\rm rel}$.

The magnetar's measured transverse velocity $v_{\rm tr}\approx240~{\rm
  km/s}$ \citep{bower15} suggests it only travels
$\gtrsim3\times10^5~{\rm km}$ (0.002 astronomical units; AU) in
20~minutes.  If the observed decrease in 44~GHz flux resulted from a
decrease in ambient density, the magnetar would have had to pass
through a region with an extremely steep density gradient of $\nabla
\rho = 5\Delta n_{5} \times10^7~{\rm cm^{-3}~AU^{-1}}$, where $\Delta
n_5 \equiv \Delta n/10^5~{\rm cm}^{-3}$ (with $n \approx 10^5~{\rm
  cm}^{-3}$ the typical density cited by \citet{yusef15}) -- a highly
unlikely event.  Therefore, a sharp decrease in unpulsed 41~GHz flux
would instead come from a decrease in $v_{\rm rel}$.  Since the
magnetar's observed proper motion direction is opposite that of
blue-shifted $\sim200~{\rm km/s}$ ionized gas in the region (e.g.,
\citealt{zhao09}), $v_{\rm rel} \gtrsim 500~{\rm km/s}$ with
\citet{yusef15} suggesting that $v_{\rm rel} \sim 1000~{\rm km/s}$.
Therefore, a $\sim2$ decrease in observed flux density requires a
$v_{\rm rel}$ decrease by $\sim 700-1500~{\rm km/s}$.  Such a change
could be explained if the magnetar exited the stellar wind bubble of a
massive OB or Wolf-Rayet star.  However, such bubbles are typically
much larger than the distance traversed by this magnetar in $\sim20$
minutes.

Furthermore, if the magnetar did experience a sudden change in ambient
density of relative velocity $v_{\rm rel}$, the timescale $\Delta t$
over which the observed radio flux density will change is
approximately $\Delta t \sim R/v_{\rm rel}$, where $R$ is the radius
of the bowshock (assumed to be the size of the radio-emitting region).
For the $R\sim20$~AU as estimated by \cite{yusef15}, this suggests
that $\Delta t \sim v_{1000}^{-1}~{\rm month}$ (where $v_{\rm rel} =
1000v_{1000}~{\rm km/s}$) -- considerably larger than the $\sim20$
minute timescale measured here.  Even if changes in the radio emission
somehow occurred faster than this estimate, the $\sim170$~minutes it
would take light to traverse this $\sim20$~AU source makes it nearly
impossible for its radio emission to change by a factor of $\sim2$ in
only 20~minutes.  As a result, we conclude the high-frequency flux
density of SGR J1745$-$29 is dominated by the magnetar's pulsed
emission.  In fact, similar variability on such timescales was
observed at higher frequencies $\sim2-3$ months after our JVLA
observation on 2014 May 10 \citep{torne15}.

However, while both our analysis and that conducted by \citet{yusef15}
of JVLA observations conducted between 2013 Oct 26 and 2014 May 31
indicate that while the $41/45$ GHz flux density of this magnetar
varied significantly during this period (Table \ref{tab:ag941fluxes}
$\sim1-3$~mJy), at all epochs the flux density measured by the JVLA
was $\gtrsim5-10\times$ higher than the 45~GHz pulsed flux density of
$\sim0.1~{\rm mJy}$ measured by the GBT on 2014 April 10.  Since the
JVLA observations are sensitive to the total (pulsed and unpulsed)
radio emission from the magnetar, while the GBT observation can only
measure the magnetar's pulsed radio emission, it may be possible that
$\sim90\%$ of the magnetar's radio emission is unpulsed.  However, as
described above, the rapid factor of $\sim2$ variability in the
magnetar's total (pulsed and unpulsed) radio emission can not result
from changes in its unpulsed radio emission.  Therefore, $\gtrsim50\%$
of the magnetar's total radio emission is pulsed.  As a result,
together the GBT and JVLA observations suggest the magnetar's pulsed
high-frequency flux density can vary by almost an order of magnitude.
This conclusion is supported by earlier measurements of its pulsed
radio measurement -- e.g., an extrapolation of the the pulsed radio
spectrum measured in 2013 May by \citet{shannon13} suggest a 45~GHz
flux density of $\sim100 \mu$Jy, comparable to what we measure in our
GBT observation (\S\ref{gbt}).  Additionally the 41 GHz flux densities
measured in our analysis of service JVLA observations are consistent
with the higher frequency pulsed spectrum of the magnetar measured
between 2014 July 21 and 2014 Aug 24 \citep{torne15} -- again
supporting the notion that the magnetar primarily produces pulsed, as
opposed to unpulsed, radio emission.

\section{Summary}
\label{conclusion}

To summarize, we present measurements of the $\sim40$~GHz pulsed and
total flux density of SGR J1745$-$29 between 2013 October 26 and 2014
May 31, a period when its 8.5~GHz pulsed radio properties
significantly changed \citep{lynch14b}.  Our GBT detection of 45 GHz
pulsations on 2014 Apr 10 (\S\ref{gbt}) is the earliest detection of
$>20$~GHz pulsations from this magnetar, and the narrow, singly peaked
profile measured during this epoch is similar to the new component in
the 8.5 GHz pulsed profile which appeared a few months before
\citep{lynch14b}.  During this observation, the magnetar emitted a
single bright radio pulse in $\sim70\%$ of its rotations -- a
significantly higher fraction than previously or subsequently observed
at lower frequencies.  Furthermore, the peak flux of these bright
single 45~GHz pulses follows a log-normal distribution similar to that
measured at 8.5 GHz (Figure \ref{fig:single_fluxdist}) -- another
possible connection between the pulsed radio emission at high and low
frequencies.

Additionally, our analysis of JVLA observations of SGR J1745$-$249
during this period (\S\ref{jvla}) suggests its 41~GHz flux density
varied by $\sim5\times$, consistent with subsequent measurements of
its pulsed flux density.  The factor of $\sim2$ change in 44~GHz flux
density over a $\sim20$ minute period measured during a JVLA
observation on 2014 May 10 (Table \ref{tab:ag941fluxes}) strengthens a
magnetospheric origin for this high-frequency emission.  While
additional observations are needed to test if the connections between
the low and high frequency pulsed radio emission are real, these
results suggest that further study of this magnetar may significantly
improve our understanding of the pulsed radio emission from these
sources.

\acknowledgements
The National Radio Astronomy Observatory is a facility of the National
Science Foundation operated under cooperative agreement by Associated
Universities, Inc. 

\bibliography{ms}
\bibliographystyle{aasjournal}

\end{document}